\documentclass{article} 
\usepackage{iclr2021_conference,times}

\usepackage{amsmath,amsfonts,bm}

\def\eqref#1{equation~\ref{#1}}

\def\1{\bm{1}}

\DeclareMathAlphabet{\mathsfit}{\encodingdefault}{\sfdefault}{m}{sl}
\SetMathAlphabet{\mathsfit}{bold}{\encodingdefault}{\sfdefault}{bx}{n}

\usepackage{hyperref}
\usepackage{url}
\usepackage{graphicx}
\usepackage{CJK}
\usepackage{amsfonts}

\title{RadioNet: Transformer based Radio Map Prediction Model For Dense Urban Environments}

\author{Yu Tian, Shuai Yuan \& Naijin Liu* \\
Qian Xuesen Laboratory of Space Technology \\
China Academy of Space Technology \\
Beijing, China \\
\texttt{\{tianyu,yuanshuai,liunaijin\}@qxslab.cn} \\
\And
Weisheng Chen \\
School of Aerospace Science and Technology \\
Xidian University \\
Xi’an, China \\
\texttt{chengweisheng@xidian.edu.cn} \\
\AND
}

\begin{document}

\maketitle

\begin{abstract}
Radio Map Prediction (RMP), aiming at estimating coverage of radio wave, has been widely recognized as an enabling technology for improving radio spectrum efficiency. However, fast and reliable radio map prediction can be very challenging due to the complicated interaction between radio waves and the environment. In this paper, a novel Transformer based deep learning model termed as RadioNet is proposed for radio map prediction in urban scenarios. In addition, a novel Grid Embedding technique is proposed to substitute the original Position Embedding in Transformer to better anchor the relative position of the radiation source, destination and environment. The effectiveness of proposed method is verified on an urban radio wave propagation dataset. Compared with the SOTA model on RMP task, RadioNet reduces the validation loss by 27.3\%, improves the prediction reliability from 90.9\% to 98.9\%. The prediction speed is increased by 4 orders of magnitude, when compared with ray-tracing based method. We believe that the proposed method will be beneficial to high-efficiency wireless communication, real-time radio visualization, and even high-speed image rendering.
\end{abstract}

\section{Introduction}

With the advent of Fifth Generation (5G) wireless communication and massive growth of wireless devices, radio spectrum resources are becoming increasingly scarce [\cite{boulogeorgos2018low},\cite{rawat2015dynamic}]. Understanding the propagation behavior of radio waves is very important for making full use of limited spectrum resources [\cite{sun2018propagation},\cite{shafi2018microwave}]. Radio Map Prediction (RMP) aims at modelling how radio wave propagates in physical space and predicting radio power for every location in a geographic area, given radiation conditions of transmitter and environment geometry. Fast and reliable radio map prediction in dense urban environment has been widely recognized as an enabling technology for network planning, interference coordination, power control, spectrum management, dynamic spectrum access and beamforming prediction in 5G [\cite{van2020power}, \cite{park2019power},\cite{zhang2020radio},\cite{skidmore2016simulation}].

Fast and reliable radio map prediction is very challenging, since radio wave would reflect, refract, diffract and scatter complicatedly in environment after originated radiated from the transmitter source. Ray Tracing (RT) method is widely used in mobile communication environments to predict the propagation characteristics of radio waves [\cite{yun2018radio}, \cite{hossain2019efficient}]. However, RT-based radio map prediction suffers slow prediction speed and huge computational complexity, since all significant radiation rays arrived at all pixels in radio map need to be independently tracked.

Formally, radio map prediction is a dense prediction task, which requires dense outputs for given input. Recently, Convolutional Neural Networks (CNN) has achieved great success in fast dense prediction tasks, such semantic segmentation \cite{chen2018encoder} and depth estimation \cite{bhat2020adabins}. Instead of predicting each pixel output independently, CNN model extracts common features of for all pixel outputs leveraging the hierarchical structure and saves lots of redundant computation. On this basis, fast prediction can be achieved.

Several works have been reported on the application of CNN in radio map prediction. \cite{levie2021radiounet} introduced Unet network \cite{ronneberger2015u} for prediction radio maps given city geometry and transmitter location, which is originally used for semantic segmentation of medical images. \cite{ozyegen2020deep} proposed to use Unet model with strided convolutions and inception modules for fast radio map prediction. \cite{krijestorac2020spatial} employed Unet to interpolate radio map in urban environment using sparse signal samples collected across the prediction space and 3D map of the environment. \cite{imai2019radio} used CNN to extracted features for propagation loss prediction from spatial information. \cite{han2020power} proposed a Generative Adversarial Networks (GANs) based prediction method, in which a generator is trained to extract propagation characteristics and generate radio maps, a discriminator is trained to find flaws in generated maps and help to improve the prediction result.

Although above CNN-based approaches achieve superior prediction speed, their prediction reliability is limited. The reason is as follows. In radio map prediction, the source and destination may be far apart in space. At the same time, due to the presence of reflection and scattering, the radio wave may spread to a long distance and then come back. As a result, global information of the propagation environment is required to predict the radio power of a single pixel in radio map. However, convolution operation in CNN model cannot effective model such spatial global dependence due to its localized receptive field.

In this paper, a RadioNet model is proposed for fast and reliable radio map prediction. In particular, Transformer module is used to solve the long-range dependence modeling problem in radio wave propagation. Unlike prior CNN-based methods, Transformer module is powerful at modeling global contexts, which has been widely witnessed in the field of machine translation and natural language processing (NLP) \cite{vaswani2017attention}. However, transferring Transformer from NLP to RMP task is non-trivial. We specially designed a Transformer-based network architecture for RMP task. As far as we know, this is the first application of Transformer in the field of radio wave propagation. We show empirically that proposed method achieves superior results on zero-shot radio map prediction task. The main contributions of this paper are as follows.

\begin{itemize}	
	\item
	A novel network termed as RadioNet is proposed, in which large-scale environment feature is used to predict the average path loss, small-scale environment feature is used to predict the fine shadow effect, and Transformer layers are used at different scales to model the radio wave interaction between different regions.
\end{itemize}

\begin{itemize}	
	\item
	A novel Grid Embedding technique is proposed to substitute the original Position Embedding in Transformer, which can better anchor the relative position of the source, destination and environment and improve the generalization ability of radio map prediction.
\end{itemize}

\begin{itemize}	
	\item
	A radio map prediction dataset is developed for dense urban scene. Each sample considers a different environment setup, a random radiation location, and a random radiation frequency. This dataset puts forward a challenging zero-shot radio map prediction task, which requires model to make reliable prediction in environments it has never met.
\end{itemize}

\section{Background}

CNN models are limited to characterize the interaction between the current position and the position in its receptive field. In order to characterize interaction with other distant positions, the CNN model needs to continuously expand the radius of receptive field until the receptive field covers those positions. Due to the significant increase in computational complexity, it is difficult to use CNN to model the interaction between distant locations. Unlike CNN, Transformer directly calculates the similarity of feature embeddings between different locations. As a result, the interaction between positions can be modeled regardless of the distance between them.

\begin{figure}
	\centering
	\includegraphics[width=0.7\linewidth]{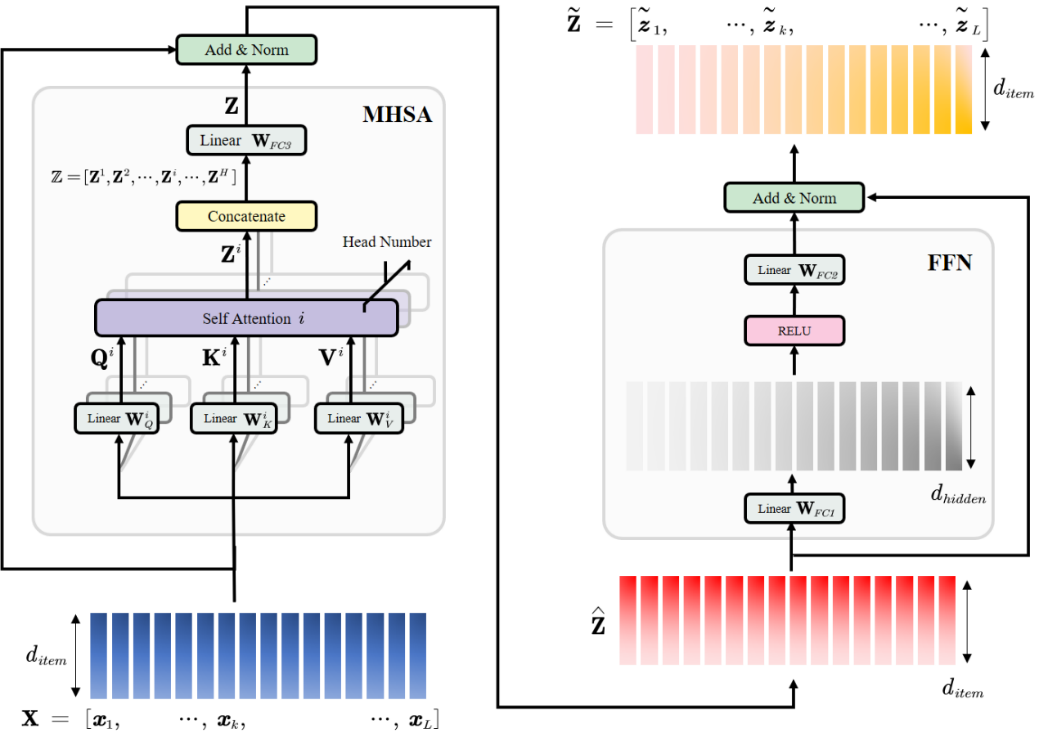}
	\caption[]{Structure of Transformer Layer}
	\label{fig:fig1}
\end{figure}

The input of Transformer Layer is an embedding of item sequence   $ \textbf{X}=[ \textbf{\textit{x}}_{1}, ..., \textbf{\textit{x}}_{k}, ..., \textbf{\textit{x}}_{L}] $, where $ \textbf{X} \in \mathbb{R} ^{d_{item}\times L}$ . $ L $  represents the length of sequence.   $ d_{item} $ represents embedding dimension. The k-th column $ \textbf{\textit{x}}_{k} $  denotes the distributional representation the k-th item in the sequence. The output of Transformer is also a item embedding sequence $  \widetilde{\textbf{Z}} \in \mathbb{R} ^{d_{item}\times L}$  . Transformer Layer aims to model the interaction between $\tilde{\textbf{\textit{z}}}_{k}$  and $ \textbf{\textit{x}}_{q}, q \in [1,L] $ , and transform the input sequence to target sequence. 

The structure of Transformer Layer is shown in Figure~\ref{fig:fig1}, which consists of a Multi-Head Self-Attention (MHSA) module and a Feed Forward Network (FFN) module. FFN is used to perform representation optimization on item embeddings. The long-range dependence learning is mainly achieved via MHSA. In MHSA, sequence embedding flows through   parallel self-attention branches, which model the correlation of input and output in different subspaces. In a particular branch $ i $, sequence embedding$  \textbf{X} $  is linearly projected to Query matrix $ \textbf{Q}^{i} \in \mathbb{R} ^{d_{item}\times L} $  , Key matrix  $ \textbf{K}^{i} \in \mathbb{R} ^{d_{item}\times L} $ , and Value matrix  $ \textbf{V}^{i} \in \mathbb{R} ^{d_{item}\times L} $ . The columns of matrix $ \textbf{Q}^{i} $  encode query operation required to predict each output item, columns of matrix $ \textbf{K}^{i} $  encode the attributes of each input item, and the columns of $ \textbf{V}^{i} $  encodes information conveyed by each input item. The multiplication between  $ \textbf{Q}^{i} $ and $ \textbf{K}^{iT} $  characterizes pairwise correlation between all outputs and all inputs. On this basis, the output is obtained by weighted summation of columns in matrix $ \textbf{V}^{i} $ , which can be written in following matrix form,
\begin{equation}\label{key}
	\textbf{Z}^{i} = softmax (\dfrac{\textbf{Q}^{i}  \cdot \textbf{K}^{iT}}{\sqrt{d_{item}}}) \cdot \textbf{V}^{i}
\end{equation}

where the superscript $ ^{T} $ is used to indicate matrix transpose. Finally, all outputs from different self-attention branches are concatenated, and a linear projection is applied to fuse all information obtained by different branches and restore the shape of sequence embedding.

\section{Method}

\begin{figure}
	\centering
	\includegraphics[width=0.7\linewidth]{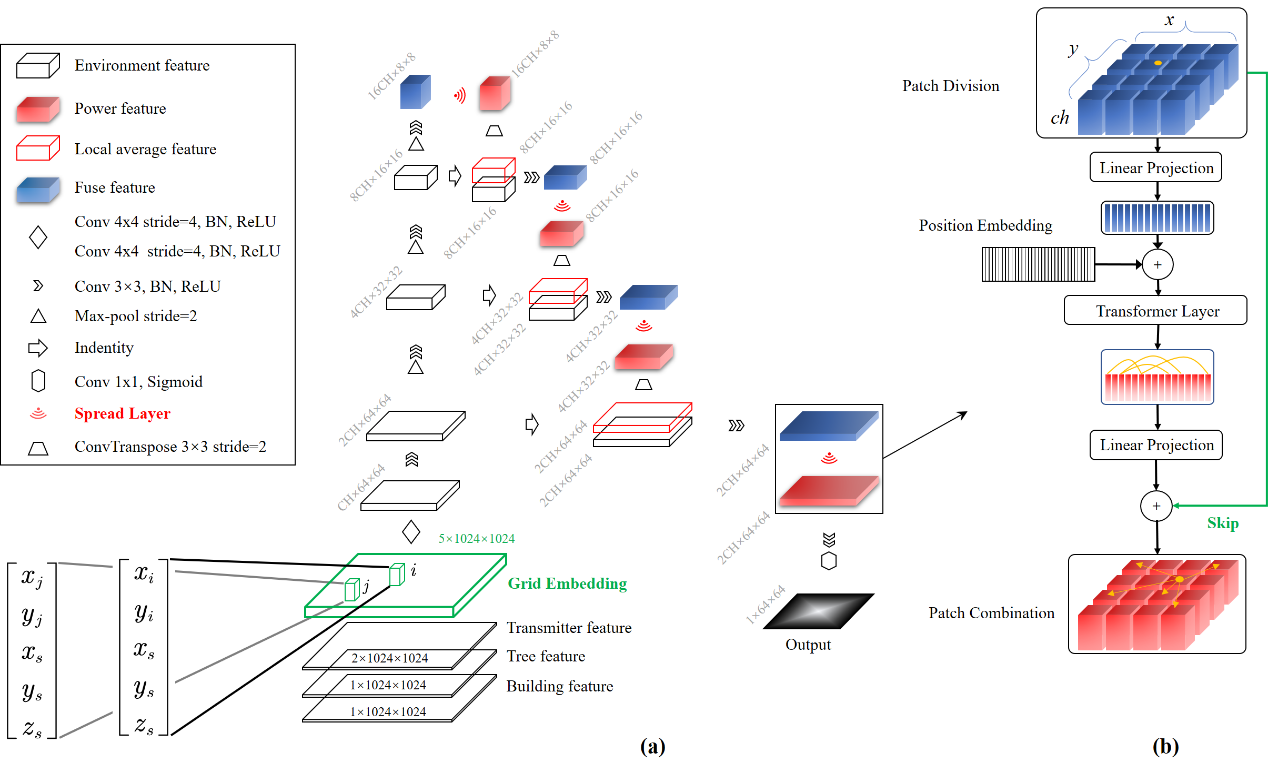}
	\caption{(a) Architecture of RadioNet. (b) Structure of spread layer.}
	\label{fig:fig2}
\end{figure}

\subsection{RadioNet Architecture}

The architecture of proposed RadioNet is illustrated in Figure \ref{fig:fig2} (a). The backbone of proposed model consists of an encoder and a decoder. The role of the encoder is to extract environment features with different scales, in which max pooling layer and convolutional layers are alternately employed. The large-scale environment feature with large receptive field is used to predict the average path loss, and the small-scale environment feature with small receptive field is used to predict the fine shadow effect.

The purpose of decoder is to interpret the radio propagation process and predict received power for all spatial locations considering the environmental characteristics and radiation source property. However, directly making high-precision pixel-level prediction is very difficult, since radio waves arriving at different spatial locations have undergone different reflection and scattering processes and it is too complex to interpret each position individually. Typically, the received powers would be similar for adjacent spatial positions, since the received power would be dominant by average path loss. The average path loss is determined by the general landscape between the source and destination. Between adjacent receiving positions, only minor differences are present, which is caused by the relative position nuances between source, destination, and obstacles in environment. On this basis, a progressive interpretation method is adopted in decoder. First, low-resolution power features are up-sampled through transposed convolution to predict the local average power field at higher resolutions. Then, a spread layer is employed to refine the prediction considering environment details around destinations and radiation source.

In order to characterize the details of radio propagation, global information needs to be considered. This is because that radio wave can travel anywhere in space before it arrives at destination. However, convolution operation is limited to model such long dependence. To solve this problem, a spread layer is proposed in this paper. 

The structure of spread layer is illustrated in Figure \ref{fig:fig2} (b). In the spread layer, 3D fused features are firstly divided equally in spatial dimensions. Then a linear layer is used to project fused features to patch embeddings. Resulted patch embedding sequence is further fed into a transformer layer in order to model the long dependency between source, destination, and environment obstacles. Finally, patch embeddings are transferred to 3D features through linear projection and patch combination.

Compared with the well-known visual transformer (ViT) model \cite{dosovitskiy2020image}, a skip connection is added between the input and output, which is more conducive to gradient back propagation and keep the path loss information in the output feature.

\subsection{Grid Embedding}

In the field of computer vision and natural language processing, Position Embedding (PE) is equipped in Transformer to anchor the location of item in sequence.
However, the PE technique may not suitable for radio map prediction task. In PE, each patch is allocated a random embedding vector to distinguish different locations, and these embedding vectors are learnable during training. This method is harmful to radio propagation deduction tasks, since the spatial relationship between the source, destination and environmental obstacles can be damaged during training. As a result, the interaction between radio wave and environment becomes unstable, which can hardly be learned by network. 

In order to overcome this problem, we proposed a novel Grid Embedding (GE) technique to more accurately describe the spatial relationship between the source, destination and environmental obstacles. In GE, grid coordinates   and transmitter location   are fed into the network, as shown in Figure \ref{fig:fig2} (a). Leveraging GE, spatial relationship remains fixed during training, and network can more easily model the radio-environment interaction.

\begin{figure}
	\centering
	\includegraphics[width=0.7\linewidth]{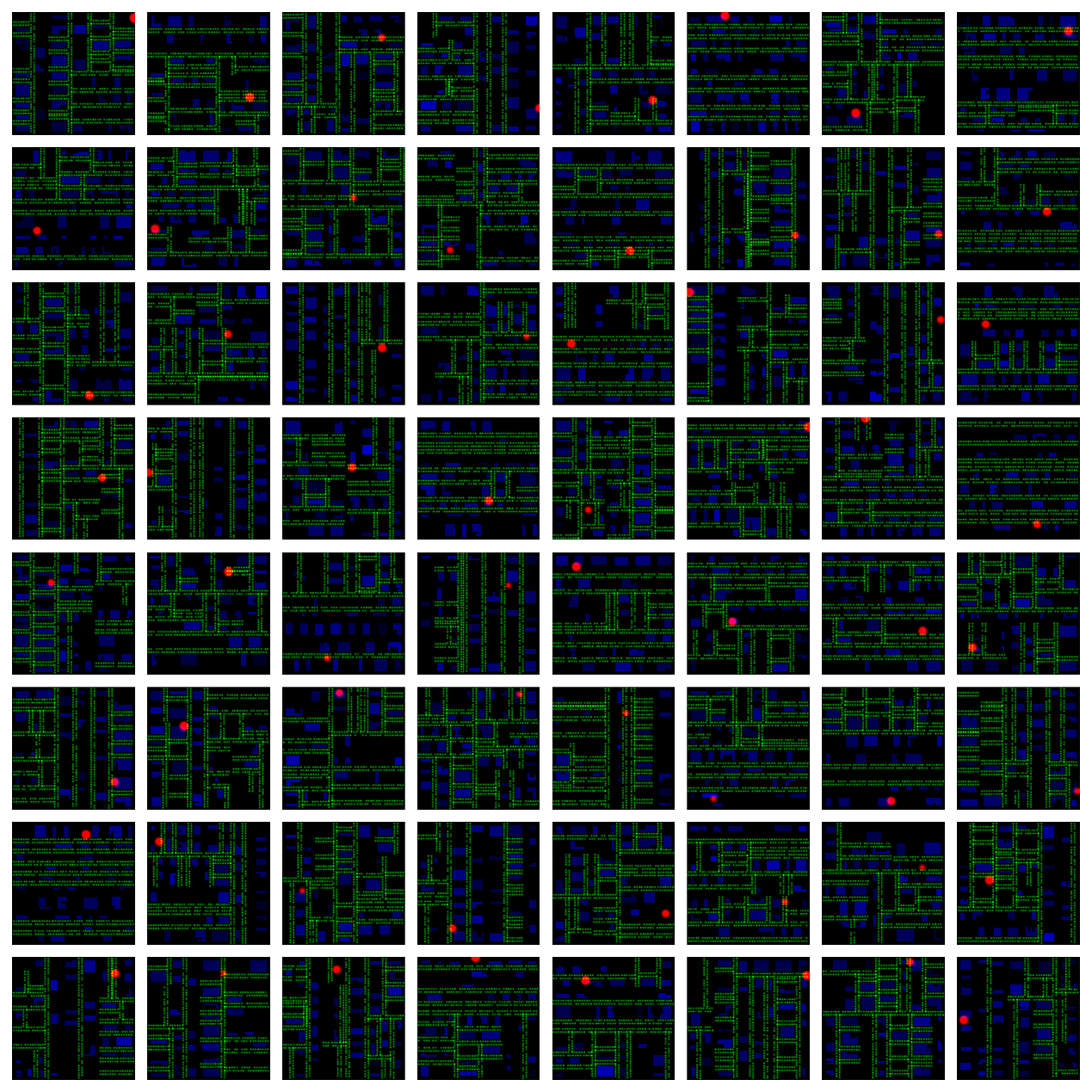}
	\caption{Environment and radiation conditions of transmitter. Colors red, green and blue are used to represent buildings, trees and radiation source. The depth of blue, depth of green and size of red point respectively indicate the height of building, tree and radiation source. Radiation frequency is not illustrated here.}
	\label{fig:fig3}
\end{figure}

\begin{figure}
	\centering
	\includegraphics[width=0.7\linewidth]{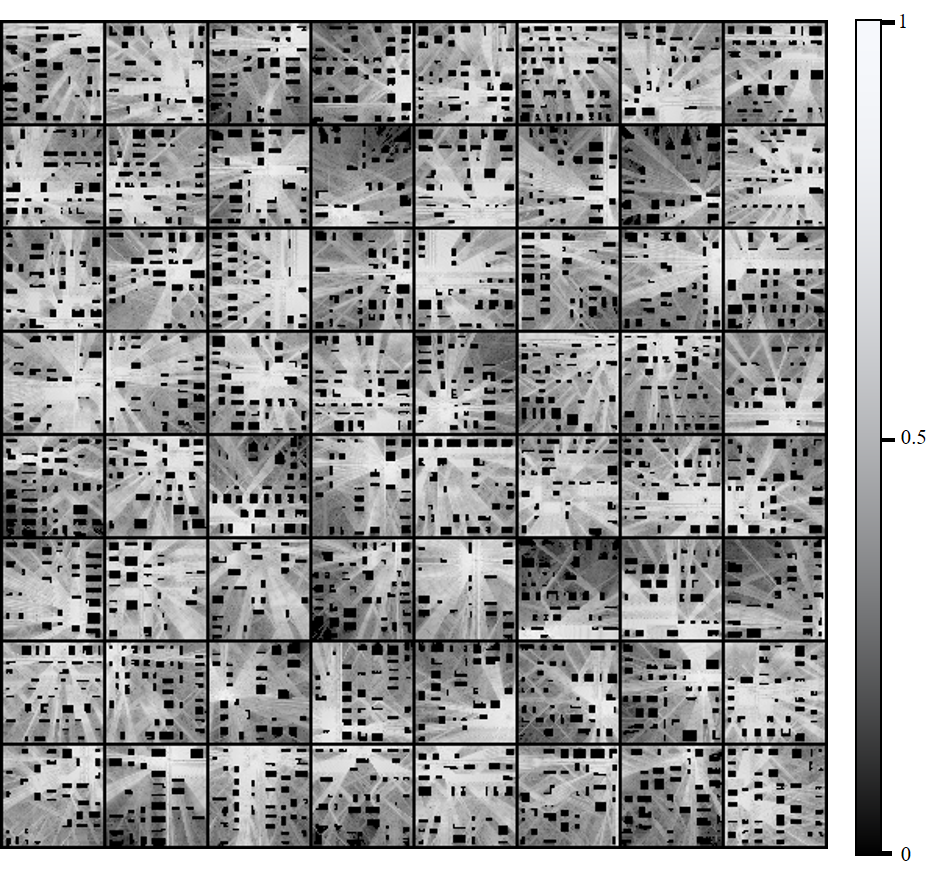}
	\caption{Ground truth. Radio maps given by ray tracing method.}
	\label{fig:fig4}
\end{figure}

\section{Urban Radio Map Dataset}
In order to verify the effectiveness of our method, a radio map dataset is developed for dense urban scenes. Each sample in the dataset considers a different environment geometry, a random radiation location, and a random radiation frequency. This dataset puts forward a challenging zero-shot radio map prediction task, which requires the model make reliable prediction in environments it has never met.

Taking the layout requirements of building, road and vegetation in typical urban scenes as constraints, environment geometries are randomly generated. Rectangular, L-shaped, n-shaped, H-shaped, E-shaped, F-shaped, Z-shaped, and T-shaped buildings are considered. The building height ranges from 30 meters to 70 meters. The vegetation is dominated by roadside trees, the average distance between trees is 10 meters, the tree height ranges from 10 meters to 50 meters. The horizontal position of the transmitter is randomly distributed in the environment, and the height ranges from 20 meters to 80 meters. The omnidirectional dipole antennas are used for radiation, and the radiation frequency ranges from 5.735GHz to 5.825 GHz.

Given radiation conditions and environment geometries, radio maps are constructed from accurate ray-tracing data. These data are obtained by Remcom Wireless InSite software, which is widely used in mmWave and massive MIMO research at both industry and academia [\cite{alkhateeb2018deep}, \cite{va2017inverse}], and has been verified with real-world channel measurements [\cite{khawaja2018indoor}, \cite{wu2016intra}]. The data set construction lasted about a month, and a total of 453825 valid samples were obtained. The batch of samples are visulized in Figure \ref{fig:fig3} and Figure \ref{fig:fig4}. The dataset will be publicly released shortly after publication.

\section{Experiment}

\subsection{Setup}

The RadioNet is implemented in Pytorch. Details of the model are shown in Figure \ref{fig:fig2} (a), in which the channel expansion factor CH is set to be 64. The resolution of input and output is set as 1024×1024 and 64×64. Higher input resolution is used to better characterize the propagation environment. In the building feature map and tree feature map, pixel value respectively represents the height of building and tree at every location. In the transmitter feature map, the first channel indicates the location of the transmitter in one-hot manner, where the nonzero pixel value represents the height of the transmitter. In the second channel, all pixels are filled with radiation frequency. The purpose is to provide frequency information for radio-environment interaction modeling for all spatial locations. All input features are normalized to [0, 1]. For all spread layers, 3D feature map is divided into 8 equal parts along the x dimension and y dimension, forming 64 patches. In Transformer Layer, $ d_{item} $ , $ H $  and  $ d_{hidden} $ are respectively set as 512, 8, and 2048. Unet is chosen as the baseline model. As far as we know, this is the state of the art (SOTA) deep learning model for RMP task. Unet used in this paper is obtained by removing the spread layers and the Grid Embedding feature in Figure \ref{fig:fig2} (a). Dataset is divided into training set and validation set according to the ratio of 99:1. The model was trained on a NVIDIA V100 32GB GPU and an Intel(R) Xeon(R) Silver 4116 CPU @ 2.10GHz. All models were trained with a batch size of 64 and using the Adam optimization algorithm with learning rate of 0.0001 for 300,000 iterations. Mean absolute loss (L1 loss) is used as the loss function. Mean square loss (L2 loss) is also tested, but no significant changes are observed.

\begin{figure}
	\centering
	\includegraphics[width=0.7\linewidth]{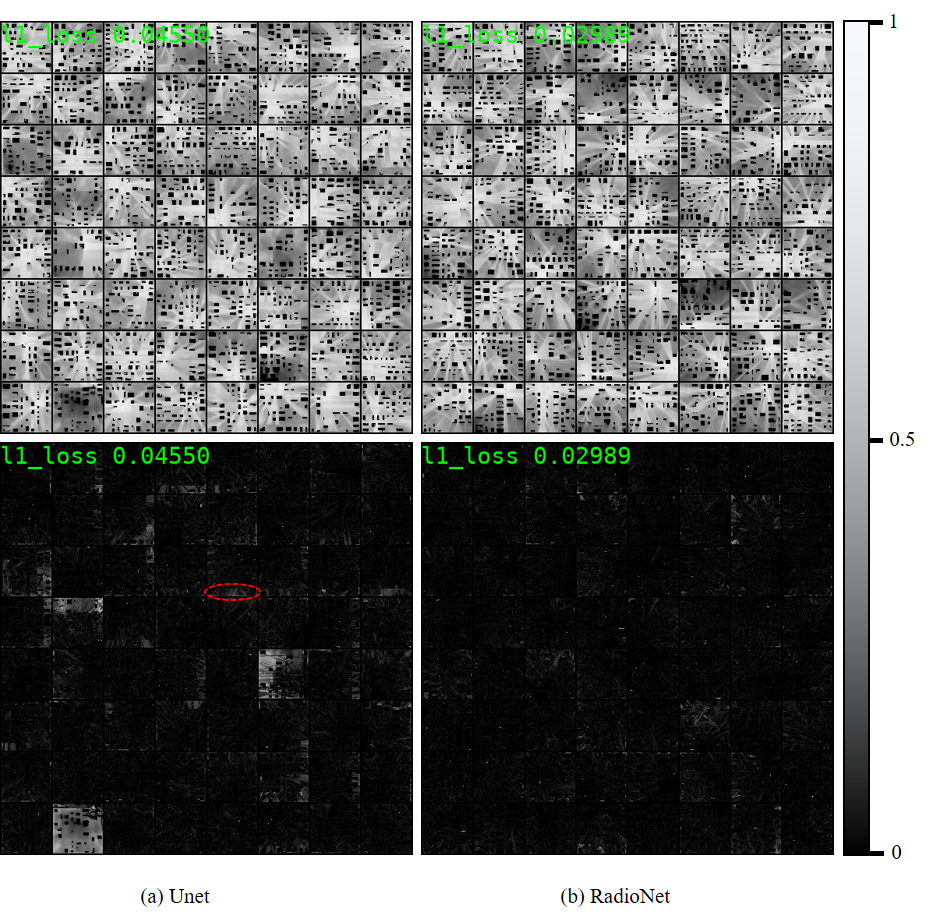}
	\caption{Prediction (first row) and error (second row) of different models. The annotations on the top left corner indicate the batch L1 loss.}
	\label{fig:fig5}
\end{figure}

\begin{figure}
	\centering
	\includegraphics[width=0.7\linewidth]{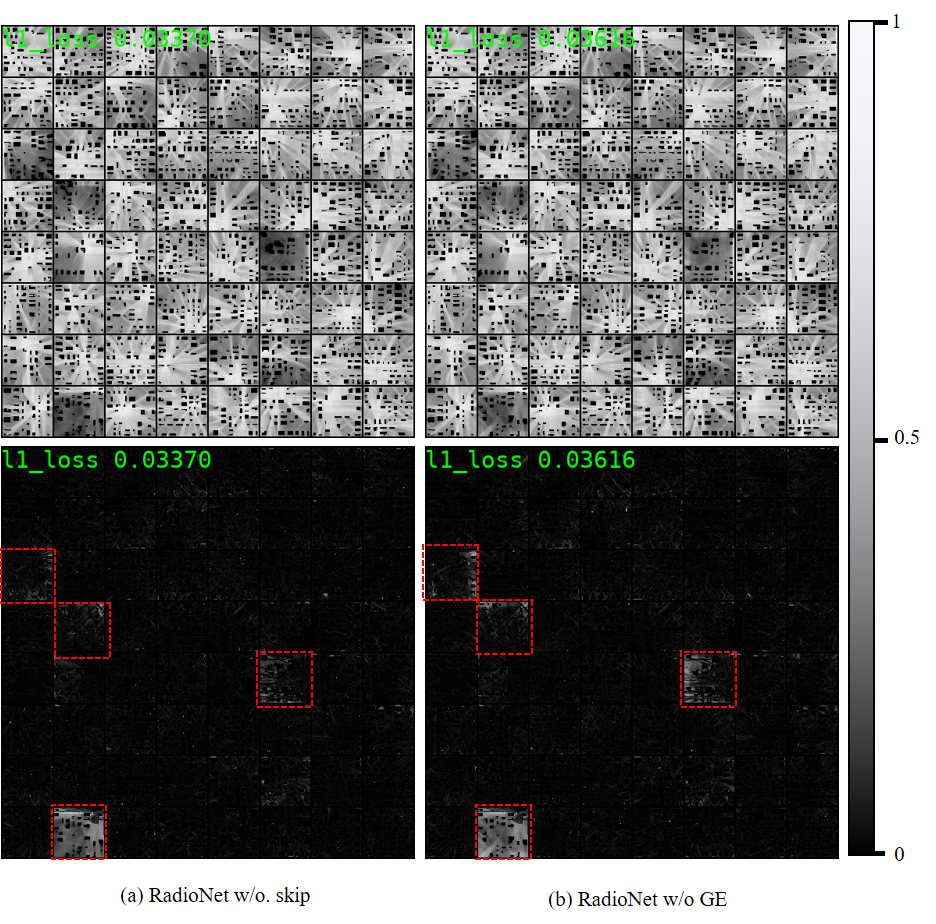}
	\caption{Effect of grid embedding and skip connection in RadioNet}
	\label{fig:fig6}
\end{figure}

\subsubsection{Results}

Prediction results on a batch in validation set are shown in Figure \ref{fig:fig5}. The first row shows the predicted radio maps. The second row shows the error between the prediction results and ground truth. Corresponding ground truth and model input are shown in Figure \ref{fig:fig4} and Figure \ref{fig:fig3}. As can be seen, there are two problems with the Unet model. Firstly, the Unet model can only give better prediction results in the area close to the radiation source. But for boundary area far away from the source, the prediction accuracy is poor, as shown in red circle in Figure \ref{fig:fig5}. The reason behind is that Unet model is limited to learn long range dependency. Secondly, Unet model fails for some input conditions, which is manifested as that the prediction errors in most spatial positions are large. These problems are largely solved by the introduction of the RadioNet architecture.

\begin{figure}
	\centering
	\includegraphics[width=0.7\linewidth]{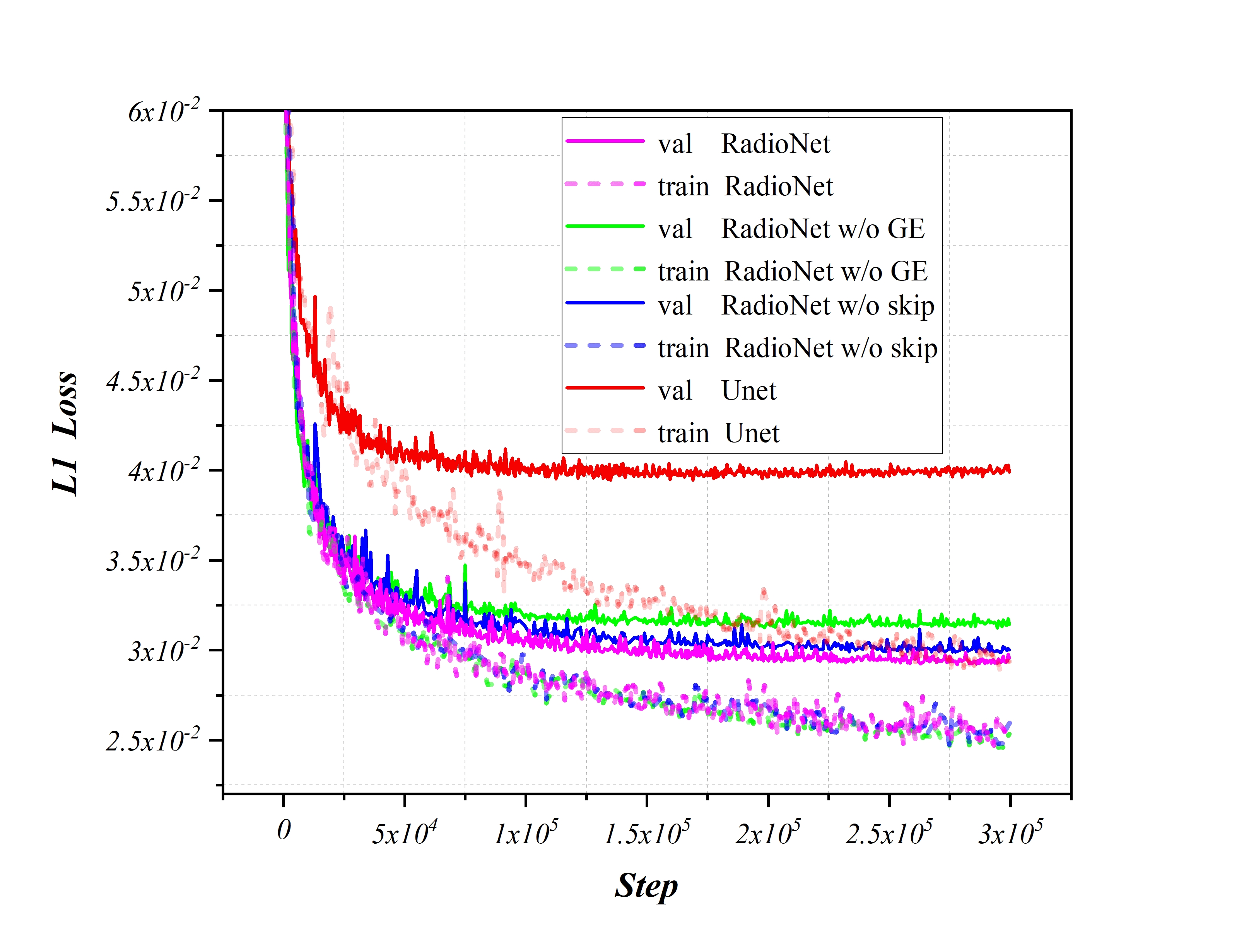}
	\caption{Loss of different models during training}
	\label{fig:fig7}
\end{figure}

The L1 loss during training is shown in Figure \ref{fig:fig7}. It can be seen that RadioNet converges faster and better than Unet. Besides, RadioNet has a lower loss on validation set, which means it has stronger generalization ability and can better model the interaction between radio waves and the environment, rather than simply memorizing specific combination of radiation and environment. In addition, it can be seen from the figure that GE and skip connection helps improve the generalization ability. The effects of removing GE and skip connection are visualized in Figure \ref{fig:fig6}. By Comparison with Figure \ref{fig:fig5} (b), it implies that combination of GE and skip connection can improve the prediction performance on difficult cases, as shown in red boxes in Figure \ref{fig:fig6} (b).

As shown in Table 1, the losses on validation set are 0.0293 for RadioNet and 0.0403 for Unet. The validation loss is reduced by 27.3\%. loss It is worth pointing out that removing PE gives similar result as RadioNet. This indicates that PE technique may not be necessary for radio map prediction, although it is required for NLP and computer vision tasks. TransUnet \cite{chen2021transunet} model is also compared here, which simply employs one Transformer layer in the bottleneck of Unet. This implies that the application of Transformer in RMP tasks is non-trivial, and careful design is indeed required to fully exploit the potential of Transformer. 

\begin{table}[t]
	\caption{Comparison of different models}
	\label{tab1}
	\begin{center}
		\begin{tabular}{ll}
			\multicolumn{1}{c}{\bf Model}  &\multicolumn{1}{c}{\bf L1 Loss on Validation Set}
			\\ \hline
			\multicolumn{1}{c}{ Unet}  &\multicolumn{1}{c}{ 0.0403} \\
			\multicolumn{1}{c}{ TransUnet}  &\multicolumn{1}{c}{0.1011} \\
			\multicolumn{1}{c}{ RadioNet w/o. skip}  &\multicolumn{1}{c}{0.0302} \\
			\multicolumn{1}{c}{ RadioNet w/o. GE}  &\multicolumn{1}{c}{0.0317} \\
			\multicolumn{1}{c}{ RadioNet w/o. PE}  &\multicolumn{1}{c}{0.0294} \\
			\multicolumn{1}{c}{ \bf RadioNet (ours)}  &\multicolumn{1}{c}{ \bf 0.0293} \\
			\hline
		\end{tabular}
	\end{center}
\end{table}

The relationship between L1 Loss l and power prediction error e is e = l*180 dB, which is determined by receiving power variation range [-250 dB, -70 dB]. We define a radio map to be effective predicted if the power prediction error is below 10 dB, and prediction reliability as the ratio of radio maps being effectively predicted. The RadioNet improves prediction reliability from 90.9\% to 98.9\% compared with Unet, which means the proposed method can significantly improve the reliability of radio map prediction.

Attributed to high reusability of low-level features and high parallelism of model architecture, the proposed RadioNet significantly improves the speed of radio map prediction. For urban scenes considered here, ray tracing method takes 66.1 seconds to predict a radio map with a resolution of 64×64. The result is obtained by commercial software Wireless Insite. In contrast, the average inference time of RadioNet is 5.5e-3 seconds. The prediction speed is increased by 4 orders of magnitude.

\section{Conclusion}

In this paper, a novel RadioNet model is proposed for fast and reliable radio map perdition, in which a Transformer based spread layer is used for model the long-range dependency of source, destination and environment. We found that Position Embedding technique originally used in computer vision and NLP is not necessary for radio map perdition task. In order to better anchor the topological relationship of transmitter source, destination and environment, a novel Grid Embedding technique is proposed. Experiment results show that our method can significantly improve the accuracy, reliability and speed of radio map perdition. In addition, we believe that our method is also suitable for other tasks that require the use of ray tracing, such as image rendering.

\bibliography{iclr2021_conference}
\bibliographystyle{iclr2021_conference}

\end{document}